\documentclass[10pt,conference,compsocconf]{IEEEtran}

\usepackage{times}

\usepackage{caption}
\usepackage{cite}
\usepackage{times}
\usepackage{xcolor}
\usepackage{tikz}
\usetikzlibrary{shapes,calc,arrows}
\usepackage{graphicx}
\usepackage{url}
\usepackage{flushend}
\usepackage{orcidlink}
\usepackage{url}

\usepackage{breakurl}
\usepackage{hyperref}
\hypersetup{hidelinks,unicode,breaklinks}

\usepackage{ifpdf}
\ifpdf
\pdfoutput=1 
\pdftrue
\fi

\makeatletter
\def\ps@IEEEtitlepagestyle{
	\def\@oddfoot{\mycopyrightnotice}
	\def\@evenfoot{}
}
\def\mycopyrightnotice{
	{\footnotesize
		\begin{minipage}{0.8\textwidth}
			\centering
			Please cite as: Michael Molle, Ulrich Raithel, Dirk Kraemer, Norbert Graß, Matthias Söllner, and Andreas Aßmuth, ``Security of Cloud Services with Low-Performance Devices in Critical Infrastructures,'' in \emph{Proc of the 10th International Conference on Cloud Computing, GRIDs, and Virtualization (Cloud~Computing~2019), Venice, Italy}, May 2019.
		\end{minipage}
	}
}
\makeatother
\makeatletter
\let\blx@rerun@biber\relax
\makeatother

\DeclareRobustCommand*{\IEEEauthorrefmark}[1]{%
	\raisebox{0pt}[0pt][0pt]{\textsuperscript{\footnotesize #1}}%
}

\newcommand{\raute}[1]{
	\draw[fill=black] #1 -- ++(-.15,0)-- ++(.15,.15)-- ++(.15,-.15)-- cycle;
	\draw[fill=white] #1 -- ++(-.15,0)-- ++(.15,-.15)-- ++(.15,.15)-- cycle;
	
}

\begin{document}
\pagenumbering{gobble}

\title{\textbf{\Large Security of Cloud Services with Low-Performance Devices in Critical Infrastructures}\\[0.2ex]}

\author{%
	\IEEEauthorblockN{~\\[-0.4ex]\large Michael Molle\IEEEauthorrefmark{1}, Ulrich Raithel\IEEEauthorrefmark{1}, Dirk Kraemer\IEEEauthorrefmark{2}, Norbert Gra\ss\IEEEauthorrefmark{3}, Matthias Söllner\IEEEauthorrefmark{4}, and Andreas Aßmuth\IEEEauthorrefmark{4}\,\orcidlink{0009-0002-2081-2455}\\[0.3ex]\normalsize}
	\IEEEauthorblockA{\IEEEauthorrefmark{1}SIPOS Aktorik GmbH, Altdorf, Germany, Email: {\tt \{michael.molle $|$ ulrich.raithel\}@sipos.de}}
	\IEEEauthorblockA{\IEEEauthorrefmark{2}AUMA Riester GmbH \& Co. KG, M\"ullheim, Germany, Email: {\tt dirk.kraemer@auma.com}}
	\IEEEauthorblockA{\IEEEauthorrefmark{3}Grass Power Electronics GmbH, Nuremberg, Germany, Email: {\tt norbert.grass@grass-pe.com}}
	\IEEEauthorblockA{\IEEEauthorrefmark{4}Technical University of Applied Sciences OTH Amberg-Weiden, Amberg, Germany,\\Email: {\tt \{m.soellner $|$ a.assmuth\}@oth-aw.de}\\[1ex]}
}

\maketitle

\begin{abstract}
As part of the Internet of Things (IoT) and Industry~4.0 Cloud services are increasingly interacting with low-performance devices that are used in automation. This results in security issues that will be presented in this paper. Particular attention is paid to so-called critical infrastructures. The authors intend to work on the addressed security challenges as part of a funded research project, using electrical actuators and battery storages as specific applications. The core ideas of this research project are also presented in this paper.
\end{abstract}


\IEEEpeerreviewmaketitle

\section{Introduction}
The increasing integration of the Internet of Things into industrial production has lead to the next industrial revolution called ``Industry~4.0''. \cite{industry40} Increasing digitisation and automation leads to a greater number of systems being connected to the Cloud. This also means that in addition to traditional IT systems a growing number of Operational Technology (OT) systems is also connected to Cloud services. Nowadays, even Supervisory Control And Data Acquisition (SCADA) systems without a suitable built-in Industry 4.0 implementation will be hard to find. All of this leads to the so-called ``Industrial Internet of Things" (IIoT) as a part of the IoT.\par 
However, besides the big SCADA systems there is a great variety of embedded systems on devices like sensors, storage systems and actors running in physical processes. A power plant, for example, has only one process control system, but a couple of thousands of actuators to control the actual processes of energy generation. In recent years, many of these devices have been connected to Cloud services for advanced analytics that cannot be computed on the devices themselves because of their limited resources concerning computing power or memory. These embedded devices very often consist of a low-cost micro controller with low clock rate (usually in double-digit MHz range), using proprietary protocols on proprietary operating systems, while maintaining the real-time capability as topmost objective. This quite significant number of embedded devices incorporates a steadily growing part of the processes and infrastructure of whole branches of industrial production. It also means that industry and economy of whole countries more and more rely on such components.\par 
The government of each individual country defines for itself which processes and infrastructures are especially important and which sectors of infrastructure have to be considered critical. In Germany, for instance, these critical infrastructures are devided into nine sectors, namely energy supply, information technology and communication, transportation and traffic, health, water supply and wastewater disposal, food provisions, finance and insurance industry, government and administration, and, finally, media and culture. \cite{critis-de} In the United States of America, a similar definition comprises even sixteen critical sectors. \cite{critis-us} 
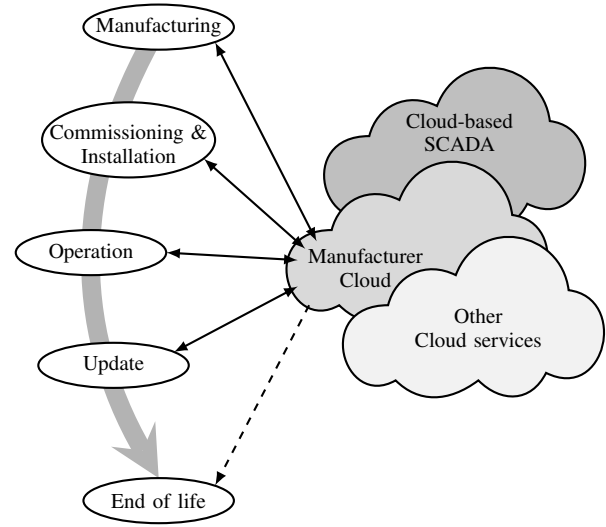
\begin{figure}[htbp]
	\centering%
	\footnotesize
	\begin{tikzpicture}
	
		\newcommand{\cloud}[3]{%
			\draw[thick, fill=#3] (#1-1.6,#2-0.7) .. controls (#1-2.3,#2-1.1)
				and (#1-2.7,#2+0.3) .. (#1-1.7,#2+0.3) .. controls (#1-1.6,#2+0.7)
				and (#1-1.2,#2+0.9) .. (#1-0.8,#2+0.7) .. controls (#1-0.5,#2+1.5)
				and (#1+0.6,#2+1.3) .. (#1+0.7,#2+0.5) .. controls (#1+1.5,#2+0.4)
				and (#1+1.2,#2-1) .. (#1+0.4,#2-0.6) .. controls (#1+0.2,#2-1)
				and (#1-0.2,#2-1) .. (#1-0.5,#2-0.7) .. controls (#1-0.9,#2-1)
				and (#1-1.3,#2-1) .. cycle;
		}
	
		\begin{scope}[draw=black!30, fill=black!30]
			\draw[line width=2.5mm, -stealth] (0, 0) to[bend right=30] (0, -6);
		\end{scope}
		\draw[thick, fill=white] (0, 0) ellipse (1cm and 0.3cm);
		\node at (0, 0) {Manufacturing};
		\draw[thick, fill=white] (-0.4, -1.5) ellipse (1.15cm and 0.5cm);
		\node[text width=2.25cm, text centered] at (-0.4, -1.55) {Commissioning \&\\Installation};
		\draw[thick, fill=white] (-0.9, -3) ellipse (1cm and 0.3cm);
		\node at (-0.9, -3) {Operation};
		\draw[thick, fill=white] (-0.6, -4.5) ellipse (1cm and 0.3cm);
		\node at (-0.6, -4.5) {Update};
		\draw[thick, fill=white] (0, -6.3) ellipse (1cm and 0.3cm);
		\node at (0, -6.3) {End of life};
		
		\cloud{4.5}{-1.75}{black!25}
		\node[text width=2cm, text centered] at (4, -1.4) {Cloud-based SCADA};
		\cloud{4}{-3}{black!15}
		\node[text width=2cm, text centered] at (2.75, -3.2) {Manufacturer Cloud};
		\cloud{4.75}{-4}{black!5}
		\node[text width=2cm, text centered] at (4.25, -4) {Other\\Cloud services};
		
		\draw[thick, latex-latex] (0.75, -0.2) -- (2.1, -2.85);
		\draw[thick, latex-latex] (0.6, -1.75) -- (1.95, -2.95);
		\draw[thick, latex-latex] (0.1, -3) -- (1.85, -3.1);
		\draw[thick, latex-latex] (0.2, -4.325) -- (1.85, -3.45);
		\draw[thick, dashed, -latex] (2, -3.7) -- (0.75, -6.1);
	\end{tikzpicture}
	\setlength{\belowcaptionskip}{-9pt}
	\caption{Lifecycle of a low-performance device and its connection\\ to Cloud services. \cite{embedded2019}}\label{fig:lifecycle}
\end{figure}
Because of the high security requirements for critical infrastructures, not only the operation of such a low-performance device must be taken into account, but all cross-relationships to Cloud services that occur during the life cycle of the device must be considered, too (cf. Figure~\ref{fig:lifecycle}). The manufacturer of the low-performance device stores specifications or maybe even initial versions of the device's firmware in their Cloud. When the device is installed in an industrial plant, it needs to be commissioned in order to communicate with the manufacturer's Cloud service. During operation, the device communicates with the Cloud service. It sends, for example, sensor data that is analysed maybe not only by the manufacturer, but also by one of the already mentioned Cloud-based SCADA systems. Therefore, an interface or gateway is needed to interconnect the manufacturer's Cloud service and the Cloud-based SCADA system. It can not be ruled out that the data is shared with other Cloud services, too. Because of known security issues or in case of new additional features, there might be updates for the software of the device. At the end of the lifecycle, e.g., when the device is broken or it no longer meets the requirements and therefore needs to be replaced, the manufacturer may wish to swipe all data and zeroise the device.\par 
The paper is structured as follows: in Section~II, we discuss threats and security challenges for Cloud-based SCADA systems as well as connected operational technology devices. In Section~III, we review related work and present our own approach in Section~IV. This approach is the subject of a current grant proposal by the authors, the different project partners are named in Section~V. We conclude in Section~VI with an outlook on future work.

\section{Threats and security challenges}
Most countries consider energy and water supply as critical sectors deserving special protection~-- and the increasing number of cyberattacks~\cite{bsi-lage} confirm this assessment to be correct. In recent years, there have been numerous attacks, like the Ukrainian blackout in 2015, when 225,000 people were suffering for a number of hours from a power outage. \cite{blackout-ukraine} During this attack, not only Industrial Control System (ICS) but also the firmware of serial-to-Ethernet adapters was damaged in order to disconnect servers from their Uninterruptible Power Supply (UPS) to maximise the length of the blackout. In December 2016, there was another attack on the Ukrainian energy supply which again resulted in a blackout for 100,000 to 200,000 people over a period of several hours. \cite{blackout-ukraine2} Such targeted attacks are no longer carried out by single attackers but by full groups with considerably different motivations. It is likely that groups of organised crime or intelligence services might be involved.\par 
The lifecycle of such an embedded device used in critical infrastructures, as described above and depicted in Figure~\ref{fig:lifecycle}, can be used to identify many attack vectors. If the adversary has access to the manufacturer's Cloud service, he could attempt to install backdoors in the initial firmware while the device is being manufactured. In addition to that, the specifications stored in the Cloud would surely be interesting for competitors and also be helpful to the attacker to detect vulnerabilities that can be exploited later. During the on-site installation, an attacker could, in principle, redirect the connection to the manufacturer's cloud service via a computer controlled by him as a starting point for a man in the middle attack. Security issues during operation are discussed explicitly in the following sections. Based on the last two phases, ``update" and ``end of life", requirements for the protection of the manufacturer's intellectual property can be exemplified. For instance, suppose another manufacturer reproduces the embedded devices in order to sell them at a lower price. This competitor would certainly like to benefit from new features or security updates that the original device manufacturer rolls out. It must therefore be ensured that a manufacturer can distinguish their original devices from clones in order not to supply those with new firmware. Likewise, it must be ensured that at the end of an original device's lifecycle its identity cannot be copied or reused so that a cloned device can pretend to be an original one.\par 
Due to these threats some operating companies start to prevent their devices from any kind of communication to outside their own network. But most of the manufacturers, however, do not want or cannot afford to dismiss the advantages of interconnectedness, e.g., for systems like energy storages in a Smart Grid. Because this development was discernible through recent years, developments ranging from classic SCADA up to Cloud-based SCADA solutions incorporate a growing number of security-critical functions. Additionally, the corresponding norms as well as legislation were pushed along, resulting, for example, in standards like IEC~62443. Legislation in Germany also has acknowledged the problem and demands~-- in accordance with requirements for Cloud operators stated by the Federal Office of Information Security \cite{bsi-cloud} and along with a ``CE-conformity label for IT security'' for manufacturers of products for critical infrastructure applying similar rules. \cite{up-kritis-recomm}

\subsection{Security challenges for Cloud-based SCADA systems}
In recent years, numerous Cloud-based ICS or SCADA systems have been developed and are now readily available. These systems interconnect on-site low-performance operational technology devices with Cloud services that run data acquisition and data analytics algorithms. The aggregation and analysis of these huge amounts of data is then used to optimise operation of the on-premise low-performance OT devices. This means that such Cloud-based SCADA systems are vulnerable against attacks targeting their Internet connection. A Distributed Denial of Service (DDoS) attack that prevents the above mentioned data acquisition and data analytics algorithms from being available for the on-premises devices certainly affects production in a non-beneficial way. In addition, data provided to these Cloud services might cause difficulties as well because of the loop back. If a sensor is hijacked and thus its data acquisition compromised, a control system today hardly has any chance at all to determine whether the data has been manipulated or not. At best, important data is provided redundantly which usually is true in plants only if the data emitting sensors are rated as safety critical. Manipulating a seemingly unimportant measurement often bears the potential of considerably interfering with a production plant's processes. Even worse are attacks on actors controlling these processes. If, for example, one of the couple of thousands actuators in a power plant can be compromised in a way that physically perturbs the process, the shutdown of the power plant~-- and so disconnecting it from the grid in order to reach a safe condition~-- is one of the more harmless scenarios imaginable.\par 
Since OT networks benefit from having all data communication at precisely deterministic and thus predictable time slots, anomaly detection can be a means of locating interference caused by an attacker. However, direct manipulation of measurement within a sensor would not alter the sensor transmitting valid data using the proper protocol to its superior control system and anomaly detection would in most cases not recognise the data being counterfeit.\par 

\subsection{Security challenges for OT devices}
For the development of low-performance devices which are deployed in critical infrastructures, security-related topics are usually the last on the list of requirements~-- if present at all. In most cases their importance is overruled by economic concerns, since they are neither really relevant for manufacturing issues nor (at least up till now) for the customers' purchasing decisions. In addition, the following fact is also in many cases unattended: a security level for low-performance systems that is comparable to traditional IT systems can only be achieved with great effort~-- if at all possible. For economic reasons these systems' soft- and hardware is usually designed to have exactly the performance to fulfil their main purpose~-- and nothing beyond. The deployment of higher performance or more complex security procedures, with respect to small profit margins and multiply optimized supply chains, quickly leads to unprofitable and uneconomic products.\par 
Apart from such economic reasons several other factors may cause even partially secured systems to fail:
\begin{itemize}
	\item insufficient communication security,
	\item lacking authentication of communication end points,
	\item faulty implementation of algorithms,
	\item faults at the protocol level,
	\item compatibility problems with applied protocols or
	\item problems with the initial key deployment.
\end{itemize}
All this increases the probability of security breaches which are either patched only infrequently or lead to a complete replacement of these devices. \cite{up-kritis-recomm} While IT systems usually provide options to implement and install patches easily, big installations, like power plants, allow only precisely defined time slots for revisions during which systems may be patched without financial losses or penalties.

\section{Related work}
On a global scale, numerous institutions and companies are developing Cloud-based services for all kinds of devices, where they all have to consider security requirements.\par 
As an example, the GE Predix service platform connects industrial assets (such as turbines, sensors, etc.) with a Cloud in order to collect and analyse operational and historical data to allow and improve predictive maintenance. \cite{predix} An additional application security service comprises two main features: a user account and authentication service using industry standards for identity management via whitelisting (amongst others), and an access control service using policy-driven authorisation for access restriction to resources programmed in a special policy language.\par 
The AUMA Cloud is a free and secure Cloud-based solution for cost-effective asset management and predictive maintenance of AUMA actuators, promoting high plant availability. \cite{auma-cloud} It provides an easy-to-use interactive platform to collect and assess detailed device information on all the AUMA actuators in a plant. It allows plant operators to detect excessive loads or potential maintenance requirements at an early stage and take remedial action in time to prevent unexpected failures.\par 
MindSphere is an open cloud platform developed by Siemens for applications in the context of the Internet of Things. \cite{mindsphere} It stores operational data from all kinds of devices and makes it accessible through digital applications in order to allow industrial customers to make decisions based on factual information. Assets can be securely connected to MindSphere with auxiliary products (e.g., MindConnect IoT2040 or MindConnect Nano) that collect and transfer relevant machine and plant data.

\section{The iSEC approach}
The authors have submitted a funding proposal entitled ``Intelligent Security for Electric Actuators and Converters in Critical Infrastructures (iSEC)" in order to solve some of the security challenges mentioned above.\par 
The technology, which is in the scope of the authors of this paper, like actuators from SIPOS and battery storage combined with electric vehicle chargers from GPE, belongs to such critical infrastructure due to the widely distributed type of the installation and remote operation of such systems. The idea behind the funding proposal is to develop an integrated data communication which facilitates both, a high internal computing performance for the processing of real-time control algorithms and secured  communication.\par  
Primarily, the untampered local operation of the equipment needs to be ensured at any time and therefore the local firmware needs to be secured from any unauthorised access. Additionally, the local equipment's data communication containing real-time signals to system wide controllers or Cloud services is essential for proper and stable plant or grid operation. For service purposes, local equipment needs to be accessible by service staff to integrate new features into the system. The confidentiality of data and signals needs to be considered and ensured.\par 
As stated before, microcontroller-based systems usually provide only very limited computing power and memory. Because of that, the computation of state of the art cryptographic algorithms or key negotiation algorithms may take several minutes. Almost all of these systems are run in environments where real-time requirements demand response times in the range of milliseconds or even microseconds, e.g., frequency converters in energy smart grids. Thus, system performance represents a significant limitation to the effectiveness of cryptographic operations. A further limitation is restricted amount of system memory~-- cryptographic algorithms have to be tailored to fit into the available RAM and ROM. As an approach to solve this problem the research of ``lightweight cryptography'' for low-performance embedded systems is just at its beginnings. \cite{lwc}\par 
Energy storage systems in larger quantities are essential to integrate higher contents of renewable energy sources into public distribution grids. Fluctuating power generation of photovoltaic or wind power systems requires short term storage to match the exact value of power consumption at any time of the day. Stationary energy storage systems and electric vehicle chargers become more common and are currently being installed into industrial buildings which are connected to public distribution grids. With increasing numbers, storage devices contribute to grid stability and therefore, they become critical infrastructure for grid operation and grid reliability. Data security becomes an important issue, as these systems are equipped with fully digital control systems, which are connected to remote systems for control and service  access functionality. Furthermore, firmware updates can be installed via remote access, which is a very useful and system-critical feature likewise. Therefore, such critical systems need to be able to verify the data they receive and to authenticate the sender of the data before starting any actions based on the data received. Additionally, the data requires confidentiality to protect the systems from competitors and invaders.\par 
\begin{figure}[htbp]
	\centering%
	\begin{tikzpicture}
	\footnotesize

	\draw[fill=yellow!20!white,draw=yellow!20!white] (0,8.1) rectangle (8,10);
	\draw (.9,9.6)node[draw](){\bfseries Level 3:};
	
	\draw[blue,thick] (2.5,9.5) -- ++(0,-.8);
	\draw[blue,thick] (3.5,9.5) -- ++(0,-.8);
	\draw[blue,thick] (4.5,9.5) -- ++(0,-.8);
	\draw[blue,thick] (5.5,9.5) -- ++(0,-.8);
	\draw[blue,thick] (7,9.5) -- ++(0,-.8);
	
	\draw(2.5,9.5) node(){\includegraphics[height=.8cm]{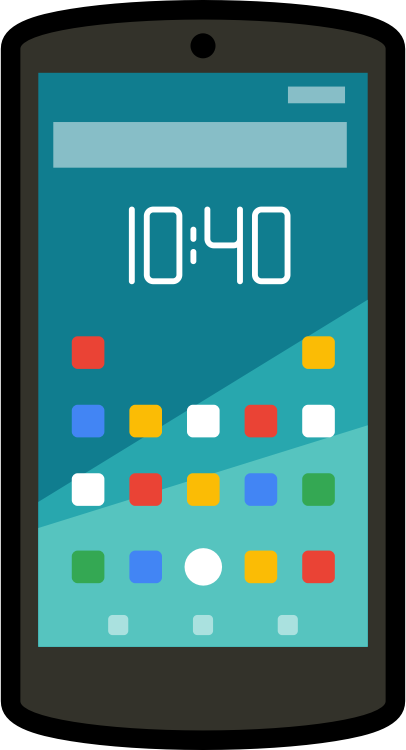}};
	\draw(3.5,9.5) node(){\includegraphics[height=.8cm]{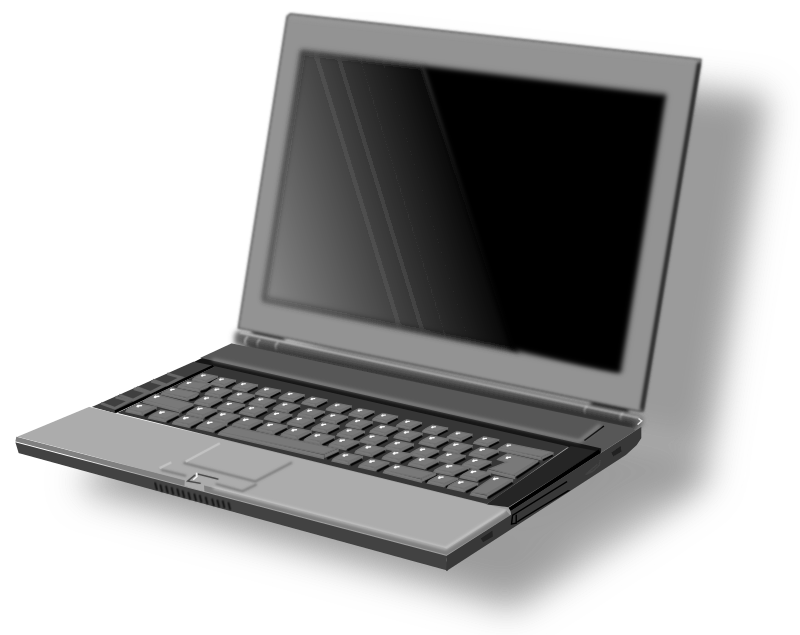}};
	\draw(4.5,9.5) node(){\includegraphics[height=.6cm]{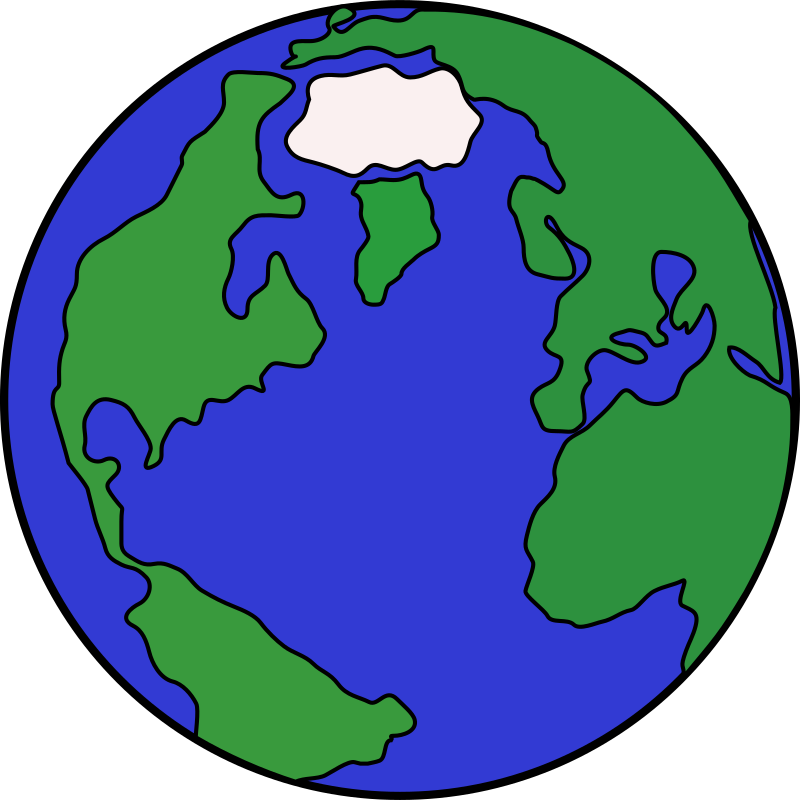}};
	\draw(5.5,9.5) node(){\includegraphics[height=.8cm]{nb.png}};
	\draw(7,9.5) node(){\includegraphics[height=.6cm]{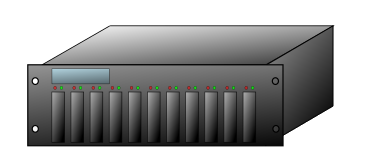}};
	
	\draw  (7.4,9.2) node[cylinder, shape border rotate=90, draw,minimum height=.25cm,minimum width=.4cm,thick,fill=black!30] (){};
	
	\draw[fill=black!20!white,draw=black!20!white] (0,3) rectangle (8,8.1);
	
	\draw[blue,thick] (5.7,8.7) -- ++(0,-.8);
	
	\draw  (4.5,8.7) node[cylinder, shape border rotate=0, draw,minimum height=5.75cm,minimum width=.1cm,thick,blue, fill=blue!20] (){};
	
	\draw  (4.5,8.68) node[] (){\tiny TCP/IP};
	
	\draw[fill=green!20!white,draw=black,thick] (0.2,3.3) rectangle (7.8,7.8);
	\draw (1.1,7.4)node[draw](){\bfseries Level 2:};
	
	\draw (3.9,7.5)node[](){\tiny\bfseries Universal Communication Server (UCS)};
	
	\draw  (1.25,5.35) node[fill=white,draw,minimum height=3.3cm, minimum width=1.8cm] (){\begin{minipage}{1.5cm}\begin{center}
		\footnotesize Application 
		\end{center}\end{minipage}};
	
	\draw[fill=blue!20!white] (2.3,6.3) rectangle (7.8,7.4); 
	
	\draw  (3.5,6.9) node[fill=white,draw,minimum height=.8cm, minimum width=.8cm] (){\begin{minipage}{.8cm}\begin{center}
		\tiny Modbus \\ TCP/IP
		\end{center}\end{minipage}};

	\draw  (4.6,6.9) node[fill=white,draw,minimum height=.8cm, minimum width=.8cm] (){\begin{minipage}{.8cm}\begin{center}
		\tiny Web-\ \ \ \\Server
		\end{center}\end{minipage}};
	\draw(4.9,7.1) node(){\includegraphics[height=.2cm]{world.png}};
	
	\draw  (5.7,6.9) node[fill=white,draw,minimum height=.8cm, minimum width=.8cm] (){\begin{minipage}{.8cm}\begin{center}
		\tiny \ \\ \ \\ Obj.Prot
		\end{center}\end{minipage}};
	
	\draw(5.7,7.05) node(){\includegraphics[width=.8cm]{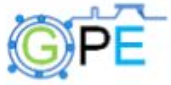}};
	
	\draw[fill=black!10](2.3,3.7) rectangle (7.8,6.3);
	\draw[fill=black!25](2.3,5.4) rectangle (3.5,6.3);
	\draw (2.9,6.02)node (){\tiny Config.XML};

	\draw(2.7,5.7) node(){\includegraphics[width=.3cm]{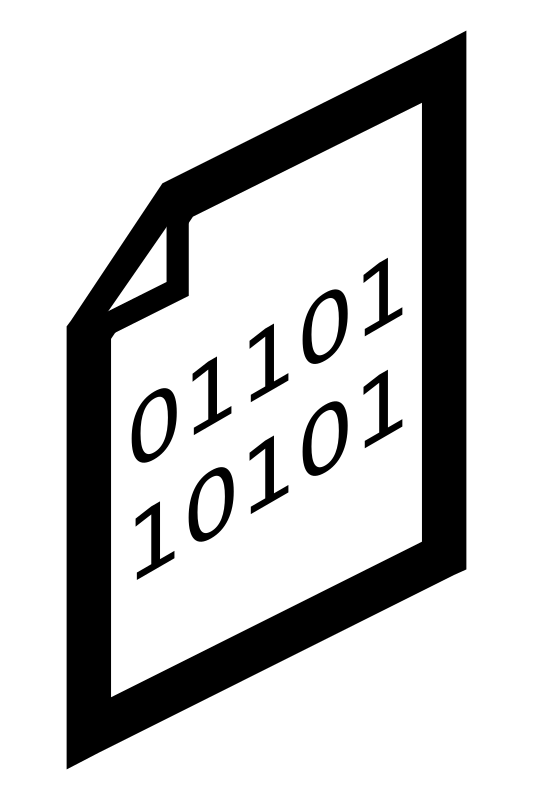}};
	\draw(3.2,5.7) node(){\includegraphics[width=.3cm]{file.png}};
	
	\draw (6.9,5.3) -| (5.1,5.8);

	\draw  (4.8,6.00) node[fill=black!0,draw,minimum height=.45cm, minimum width=.8cm] (){};
	
	\draw  (4.8,6) node[] (){\begin{minipage}{.5cm}\begin{center}
		\tiny Signal\\ Matrix
		\end{center}\end{minipage}};

	\raute{(2.8,6.3)}
	\raute{(5.7,6.3)}
	
	\raute{(5.7,7.8)}
	
	\draw[fill=yellow!70] (2.8,3.7) rectangle (7.8,5.0);
	
	\draw[latex-latex,thick] (4,5.15) |- ++(.4,.75);
	\draw[latex-latex,thick] (4.1,5.05) -- ++(.15,.15) -- ++(.3,0) -| ++(.2,.57);
	\draw[-latex,thick] (3.2,5) -- ++(0,.4);
	
	\draw(4.0,5.6) node(){\includegraphics[height=.3cm]{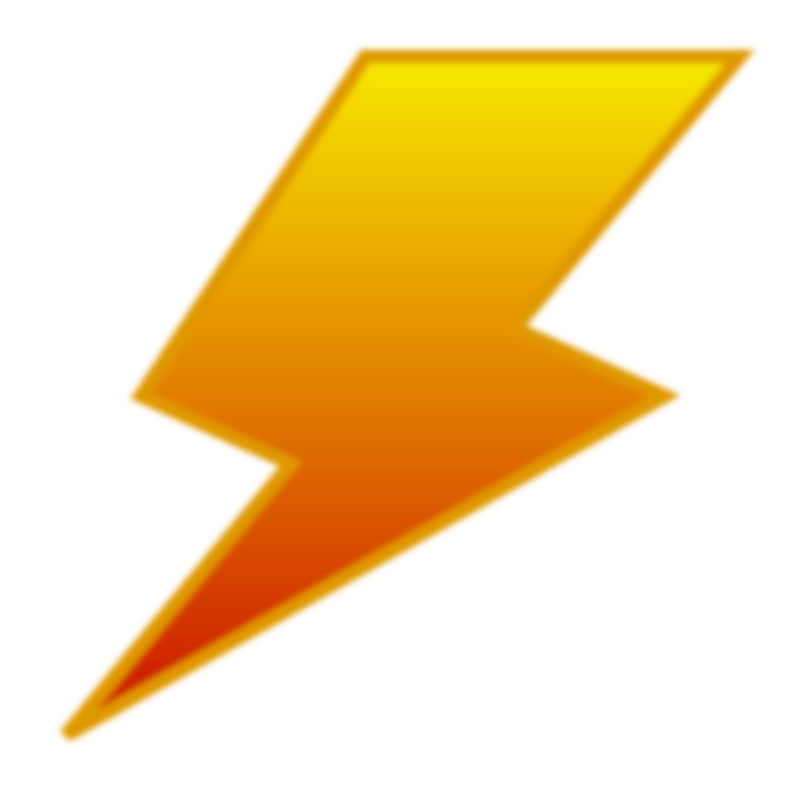}};
	\draw(4.75,5.4) node(){\includegraphics[height=.3cm]{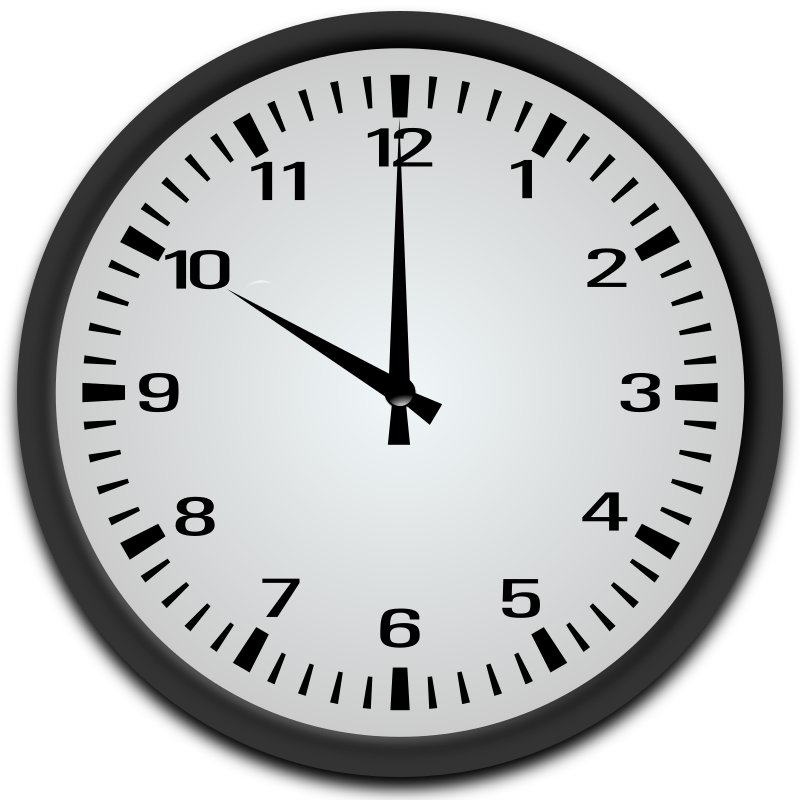}};
	\draw (5.05,5.1) node (){\tiny Timer};
	\draw (4.25,5.45) node (){\tiny Event};
	
	\draw(5.2,5.85)-- ++(.3,0);
	
	\draw(5.7,5.85)-- ++(.3,.3);
	\draw(5.7,5.95)-- ++(.3,-.3);

	\draw[fill=white] (5.4,6) rectangle ++(.4,-.3);
	\draw[fill=blue] (5.4,6) rectangle ++(.4,-.05);
	\draw[fill=white] (5.9,6.2) rectangle ++(.3,-.2);
	\draw[fill=blue] (5.9,6.2) rectangle ++(.3,-.05);
	\draw[fill=white] (5.95,5.8) rectangle ++(.3,-.2);
	\draw[fill=blue] (5.95,5.8) rectangle ++(.3,-.05);
	
	\draw  (5.8,5.4) node[] (){\begin{minipage}{1.5cm}\begin{center}
		\tiny Data   Objects
		\end{center}\end{minipage}};
	
	\draw  (7.3,5.35) node[] (){\begin{minipage}{.5cm}\begin{center}
		\tiny Data\\ Log
		\end{center}\end{minipage}};
	
	\draw  (7.3,5.4) node[cylinder, shape border rotate=90, draw,minimum height=.75cm,minimum width=.8cm,thick] (){};
	
	\raute{(3.2,5)}
	\raute{(4,5)}
	
	\draw[fill=black] (2.3,5) -- ++(0,-.15)-- ++(.15,.15)-- ++(-.15,0.15)-- cycle;
	\draw[fill=white] (2.3,5) -- ++(0,-.15)-- ++(-.15,.15)-- ++(+.15,0.15)-- cycle;
	
	\draw[fill=orange!40!white,draw=orange!20!white] (0,0) rectangle (8,3);
	\draw (.9,2.6)node[draw](){\bfseries Level 1:};
	
	\draw[thick](3.3,4) -- ++ (0,-.8);
	\draw[thick](4,4) -- ++ (0,-.8);
	\draw[thick](4.7,4) -- ++ (0,-.8);
	\draw[thick](5.4,4) -- ++ (0,-.8);
	\draw[thick](6.1,4) -- ++ (0,-.8);
	\draw  (3.3,4.25) node[fill=white,draw,minimum height=.9cm, minimum width=.6cm] (){\begin{minipage}{.4cm}\begin{center}
		\tiny CAN
		\end{center}\end{minipage}};
	
	\draw  (4,4.25) node[fill=white,draw,minimum height=.9cm, minimum width=.6cm] (){\begin{minipage}{.4cm}\begin{center}
		\tiny \ \\ \ \\  CAN
		\end{center}\end{minipage}};
	
	\draw(4,4.4) node(){\includegraphics[width=.6cm]{GPE.png}};
	\draw  (4.7,4.25) node[fill=white,draw,minimum height=.9cm, minimum width=.6cm] (){\begin{minipage}{.4cm}\begin{center}
		\tiny Mod- bus RTU
		\end{center}\end{minipage}};
	
	\draw  (5.4,4.25) node[fill=white,draw,minimum height=.9cm, minimum width=.6cm] (){\begin{minipage}{.4cm}\begin{center}
		\tiny Mod- bus TCP
		\end{center}\end{minipage}};

	\draw  (6.1,4.25) node[fill=white,draw,minimum height=.9cm, minimum width=.6cm] (){\begin{minipage}{.4cm}\begin{center}
		\tiny PWM SYNC FO
		\end{center}\end{minipage}};

	\draw  (7.1,6.8) node[minimum height=.9cm, minimum width=.6cm] (){\begin{minipage}{1.4cm}\begin{center}
		Level 3\\ Interface
		\end{center}\end{minipage}};

	\draw  (7.1,4.3) node[minimum height=.9cm, minimum width=.6cm] (){\begin{minipage}{1.4cm}\begin{center}
		Level 1\\ Interface
		\end{center}\end{minipage}};

	\draw[green!70!black, ultra thick] (3,1.7)-- ++(0,.8) -| ++ (1.,.7);
	\draw[green!70!black, ultra thick] (4.7,1.7)-- ++(0,.8);
	\draw[green!70!black, ultra thick] (6.4,1.7)-- ++(0,.8)-- ++(-3,0);
	
	\raute{(3.3,3.3)}
	\raute{(4,3.3)}
	\raute{(4.7,3.3)}
	\raute{(5.4,3.3)}
	\raute{(6.1,3.3)}

	\draw (5.4,2.7)node(){\footnotesize \bfseries CAN};
	
	\draw  (3,1.7) node[fill=white,draw,minimum height=.8cm, minimum width=1.1cm] (){\begin{minipage}{1cm}\begin{center}
		\tiny DSP Controller
		\end{center}\end{minipage}};
	\draw  (3,.7) node[fill=white,draw,minimum height=1.2cm, minimum width=1.1cm] (){\begin{minipage}{1cm}\begin{center}
		\tiny Power Inverter \\ \ \\ \ \\ \ 
		\end{center}\end{minipage}};

	\draw  (4.7,1.7) node[fill=white,draw,minimum height=.8cm, minimum width=1.1cm] (){\begin{minipage}{1cm}\begin{center}
		\tiny DSP Controller
		\end{center}\end{minipage}};
	\draw  (4.7,.7) node[fill=white,draw,minimum height=1.2cm, minimum width=1.1cm] (){\begin{minipage}{1cm}\begin{center}
		\tiny Power Inverter \\ \ \\ \ \\ \ 
		\end{center}\end{minipage}};
	
	\draw  (6.4,1.7) node[fill=white,draw,minimum height=.8cm, minimum width=1.1cm] (){\begin{minipage}{1cm}\begin{center}
		\tiny DSP Controller
		\end{center}\end{minipage}};
	\draw  (6.4,.7) node[fill=white,draw,minimum height=1.2cm, minimum width=1.1cm] (){\begin{minipage}{1cm}\begin{center}
		\tiny Power Inverter \\ \ \\ \ \\ \ 
		\end{center}\end{minipage}};
	
	\draw[fill=blue!20,rounded corners=1mm] (2.7,.2) rectangle ++ (.6,.6);
	
	\draw (2.8,.3) node (A){};
	\draw (A) ++(.1,.15) -- ++(-.1,0);
	\draw (A) ++(.1,0) -- ++(0,.4);
	\draw (A) ++(.17,0) -- ++(0,.4);
	\draw (A) ++(.27,0) -- ++(0,.1)--++(-.1,.1);
	\draw (A) ++(.27,.4) -- ++(0,-.1)--++(-.1,-.1);
	
	\draw[fill=blue!20,rounded corners=1mm] (4.4,.2) rectangle ++ (.6,.6);
	
	\draw (4.5,.3) node (A){};
	\draw (A) ++(.1,.15) -- ++(-.1,0);
	\draw (A) ++(.1,0) -- ++(0,.4);
	\draw (A) ++(.17,0) -- ++(0,.4);
	\draw (A) ++(.27,0) -- ++(0,.1)--++(-.1,.1);
	\draw (A) ++(.27,.4) -- ++(0,-.1)--++(-.1,-.1);
	\draw[fill=blue!20,rounded corners=1mm] (6.1,.2) rectangle ++ (.6,.6);
	\draw (6.2,.3) node (A){};
	\draw (A) ++(.1,.15) -- ++(-.1,0);
	\draw (A) ++(.1,0) -- ++(0,.4);
	\draw (A) ++(.17,0) -- ++(0,.4);
	\draw (A) ++(.27,0) -- ++(0,.1)--++(-.1,.1);
	\draw (A) ++(.27,.4) -- ++(0,-.1)--++(-.1,-.1);
	
	\draw (0,0) rectangle (8,10);
	\end{tikzpicture}
	\setlength{\belowcaptionskip}{-12pt}
	\caption{Data communication architecture. \cite{norbert-ieee}}\label{fig:gpe}
\end{figure}
Figure~\ref{fig:gpe} depicts the data communication architecture. The power converters, controlled by digital signal processors (Level~1) are connected via a local CAN network to a Linux-based system and communication controller (Level~2). The system controller has a TCP/IP interface which facilitates data communication to local or via Internet connected Level 3 devices for operation and service functionalities. While CAN communication is restricted to the local system, TCP/IP is critical as it can be accessed from outside the local system.\par 
It is planned to perform a detailed investigation of how internal and external interfaces can be constructed in a verifiable secure design, and how in-situ tests can prove their efficacy in terms of security and usability.\par 
Cloud services shall be used for mechanisms of identification and authentication, for easing the task of performing necessary software patches and thus improving facilities' outage times and service intervals.\par 
In addition, it is planned to investigate how Physical Unclonable Functions (PUF)  can  be  used  to  secure  communication between a Cloud server and (low-performance) sensor clients and to clearly identify a sensor client with a digital fingerprint. Hardware intrinsic deviations  caused  by  the  manufacturing  process  of  semiconductors can be used to identify chips \cite{puf1} and generate random encryption keys. The drawbacks of using non-volatile storage-mechanisms for storing encryption keys, can be overcome by using this relatively new approach. PUFs are a current subject of research,  different  approaches have yet been investigated. \cite{puf2}\cite{puf3} For  example,  with  arbiter  PUFs  a  race  condition  can  be generated  between  two  different  digital  paths  on  the  same semiconductor. An arbiter circuit is used to measure which of the paths won the race. With different challenges the path can be configured and for every challenge, the winner is determined. Because of the manufacturing deviations every chip will give a different response, despite having the same hardware configuration and  therefore, a  digital  fingerprint  can  be  read out.  As  the response  cannot be  read out  or  predicted  by  an  attacker  it  is called unclonable. Also, PUFs based on digital bistable storage elements,  like  SRAM  cells,  latches  or  flip  flops,  have  been demonstrated.  They  are  based  on  the  principle  of  bringing them in, in an unstable state, and letting  them  settle  in  one  of their stable states. Due to statistical variations during the manufacturing process, different chips cause different results despite the same hardware configuration. Many other solutions using deviations of  the  manufacturing  process  for  identifying  a  chip  are  conceivable. \cite{puf4} In  this  context, new  protocols  have also been investigated to secure lightweight communication based on PUFs. \cite{puf5}\cite{puf6}\cite{puf7}  Which  lightweight  PUF  based  protocols  can  be  used  for encryption  of  sensor  data  connected  to  a  cloud-server is another topic of our studies. Just recently,  first  semiconductor  devices with PUF-functionality are now readily available in order to identify hardware and  implement  a  digital  fingerprint, for example. \cite{puf8}\cite{puf9}\cite{puf10} It has to be investigated whether these semiconductor devices can be used in order to help solving some of the security challenges mentioned before.

\section{The consortium}
SIPOS Aktorik GmbH emanated in 1999 from the former actuator division of Siemens~AG in Nuremberg, since 2008 situated at Altdorf. Main proprietor of SIPOS Aktorik GmbH is the AUMA Riester GmbH~\& Co KG, Muellheim, which as a holding also provides commercial services. Today, SIPOS Aktorik GmbH employs a staff of 85 people in the departments assembly, R\&D, customer service and administration. During the last 20~years SIPOS Aktorik GmbH succeeded in positioning itself on the global market for electric actuators with an export quota of $80\,\%$. Main customers are international plant engineering and construction companies, valve manufacturers, and operating companies of conventional and nuclear power plants in Europe and Asia.\par 
Grass Power Electronics GmbH, Nuremberg, is working on grid conected stationary battery storage systems in the range of some hundreds of kilowatts. Core technology components are digital computer  modules for real time power converter control and for system control, including TCP based data communication.\par 
The security research group at the Technical University of Applied Sciences OTH Amberg-Weiden has already worked on funded research projects using lightweight cryptographic algorithms. They have also experience in developing security protocols using PUFs for authentication and device identification. \cite{ccc2016}

\section{Conclusion and Future Work}
In this paper, we showed that when it comes to combining low-performance embedded devices with Cloud services, all components must be secured to harden these systems against cyberattacks. Otherwise, compromised sensors can falsify computations and analytics performed in the cloud. And attacks against the Cloud services, e.g., a DDoS attack, has a direct impact on an ICS when it relies on a permanent connection to the cloud, too.\par 
To master the challenges of IIoT and Industry~4.0, it is imperative to consider possible vulnerabilities and attack vectors when designing such systems (``security by design'').\par 
The authors hope that their submitted grant proposal iSEC will be approved to work on these security challenges.

\end{document}